\documentclass[notitlepage,aps,letter,reprint,nofootinbib,longbibliography]{revtex4-1}

\usepackage{graphicx}
\usepackage{amsmath}
\usepackage{amssymb,physics}
\usepackage{hyperref}
\usepackage{tikz}

\usepackage[english]{babel}

\usepackage{letltxmacro}

\LetLtxMacro{\ORIGselectlanguage}{\selectlanguage}
\makeatletter
\DeclareRobustCommand{\selectlanguage}[1]{%
  \@ifundefined{alias@\string#1}
    {\ORIGselectlanguage{#1}}
    {\begingroup\edef\x{\endgroup
       \noexpand\ORIGselectlanguage{\@nameuse{alias@#1}}}\x}%
}
\newcommand{\definelanguagealias}[2]{%
  \@namedef{alias@#1}{#2}%
}
\makeatother

\definelanguagealias{en}{english}

\newcommand{\be}{\begin{eqnarray*}}
\newcommand{\ee}{\end{eqnarray*}}
\newcommand{\beq}{\begin{eqnarray}}
\newcommand{\eeq}{\end{eqnarray}}
\newcommand{\bequ}{\begin{equation}}
\newcommand{\eequ}{\end{equation}}

\newcommand{\ph}{{\phantom{\dagger}}}
\newcommand{\hc}{\mathrm{H.c.}}

\newcommand{\id}{\mathbb{I}}

\makeatletter
\newcommand{\doublewidetilde}[1]{{%
  \mathpalette\double@widetilde{#1}%
}}
\newcommand{\double@widetilde}[2]{%
  \sbox\z@{$\m@th#1\widetilde{#2}$}%
  \ht\z@=.9\ht\z@
  \widetilde{\box\z@}%
}
\makeatother

\begin{document}
\title{Chiral Flow in One-dimensional Floquet Topological Insulators}
\author{Xu Liu}
\author{Fenner Harper}
\author{Rahul Roy}
\affiliation{Mani L. Bhaumik Institute for Theoretical Physics, Department of Physics and Astronomy, University of California at Los Angeles, Los Angeles, California 90095, USA}
\date{\today}
\begin{abstract}
We propose a bulk topological invariant for one-dimensional Floquet systems with chiral symmetry which quantifies the particle transport on each sublattice during the evolution. This chiral flow is physically motivated, locally computable, and improves on existing topological invariants by being applicable to systems with disorder. We derive a bulk-edge correspondence which relates the chiral flow to the number of protected dynamical edge modes present on a boundary at the end of the evolution. In the process, we introduce two real-space edge invariants which classify the dynamical topological boundary behaviour at various points during the evolution. Our results provide the first explicit bulk-boundary correspondence for Floquet systems in this symmetry class.
\end{abstract}
\maketitle

\section{Introduction\label{sec:intro}}

Topological phases exhibit a deep connection between their edge behaviour and their properties in the bulk, a feature known as bulk-edge correspondence. As famously demonstrated in the quantum Hall effects (QHE) \cite{KlitzingNew1980,TsuiTwoDimensional1982,PrangeQuantum1990}, this correspondence means that nontrivial bulk topology is manifested at a boundary as protected edge modes, whose number and form can be predicted from quantities calculated far away from the edge. Boundary modes of this kind are robust against a wide range of perturbations, making them particularly suitable as experimental signatures. Indeed, many topological phases of free fermions, known as topological insulators and superconductors (TIs) \cite{KaneQuantum2005,FuTopological2007b,MooreTopological2007,SchnyderClassification2008,RoyTopological2009a,KitaevPeriodic2009,HasanColloquium2010,QiTopological2011}), have been detected partly through their protected edge modes \cite{KonigQuantum2007,Hsiehtopological2008,Hsiehtunable2009,HsiehObservation2009,XuObservation2014}.

In recent years, the study of topological phases has expanded to include driven systems, whose generating Hamiltonians vary periodically with time. In particular, driven systems of free fermions have been found to form dynamical analogues of TIs known as Floquet topological insulators (FTIs)\cite{KitagawaTopological2010a,JiangMajorana2011a,RudnerAnomalous2013,ThakurathiFloquet2013a,AsbothChiral2014,CarpentierTopological2015,NathanTopological2015,FruchartComplex2016,vonKeyserlingkPhase2016b,ElseClassification2016a,PotterClassification2016,RoyAbelian2016,TitumAnomalous2016a,RoyPeriodic2017}, and several of these phases have now been realised experimentally \cite{KitagawaObservation2012a,RechtsmanPhotonic2013,JotzuExperimental2014,Jimenez-GarciaTunable2015,CardanoDetection2017,MaczewskyObservation2017}. FTI phases exhibit bulk-boundary correspondence in a similar vein to their static counterparts, although the resulting edge modes can be very different: Remarkably, dynamical edge modes analogous to those of the QHE or TIs can exist even if the bulk bands are topologically trivial \cite{RudnerAnomalous2013,QuelleDriving2017}. In this way, bulk-boundary correspondences for driven systems are far richer than in static systems, and in most cases remain unexplored.

While a qualitative statement of bulk-boundary correspondence is simple to write down, proving a rigorous connection between bulk and edge properties is a more challenging endeavour. The usual procedure for deriving such a connection is to first obtain a robust invariant which describes the bulk topology, and then to relate this to a similarly robust invariant describing the edge modes. This approach and others have been used to obtain bulk-edge correspondences for a variety of static TI phases \cite{ElgartEquality2005a,EssinBulkboundary2011,FukuiBulkEdge2012,GrafBulkEdge2013,ProdanBulk2016a,KubotaControlled2017,YuBulk2017,HayashiBulkedge2017,BourneKTheoretic2017a,ShapiroBulkEdge2017,HannabussTduality2018}. In driven systems, topological invariants are well known for FTI phases with translational invariance \cite{YaoTopological2017}, and there have been studies of disordered FTI phases in some symmetry classes \cite{GannotEffects2015,TitumAnomalous2016a,RoyDisordered2016,NathanQuantized2017,GrafBulk2018}. However, while some bulk-edge correspondences have been obtained \cite{RudnerAnomalous2013,AsbothBulkboundary2013,AsbothChiral2014,Cedzichtopological2016,CedzichBulkedge2016,SadelTopological2017b,GrafBulkEdge2018,TauberEffective2018}, in most disordered cases, expressions for topological invariants remain lacking, and bulk-edge correspondences have not been attempted. 

In this paper, we obtain an explicit bulk-edge correspondence for Floquet systems belonging to Class~AIII of the Altland-Zirnbauer symmetry classification \cite{AltlandNonstandard1997,HeinznerSymmetry2005,RoyPeriodic2017} (i.e. those with a protected chiral symmetry). In the process, we define a new bulk invariant we call `chiral flow', which quantifies the topological properties of the driven system in the bulk. We complete the correspondence by relating this bulk invariant to two new edge invariants: a chiral-symmetry-breaking boundary flow midway through the drive, and a state-counting invariant at the end of the drive which determines the number of protected dynamical edge modes. The new invariants we define do not require translational symmetry and are applicable even in the presence of disorder. They reduce to existing winding number expressions \cite{FruchartComplex2016} in the translationally invariant limit.

Driven systems with chiral symmetry have been studied elsewhere in the literature, particularly in the context of quantum walks. A number of early references \cite{AsbothBulkboundary2013,AsbothChiral2014} classified one-dimensional systems in this class with translational symmetry and obtained an explicit bulk invariant.\footnote{We note, however, that the unitary loop drives we discuss in Sec.~\ref{sec:Prelim} seem to violate the invariants of Refs.~\onlinecite{AsbothBulkboundary2013,AsbothChiral2014}} A somewhat different bulk invariant was introduced and extended to higher dimensions in Ref.~\onlinecite{FruchartComplex2016}, and these phases were brought into a universal framework in Ref.~\onlinecite{FruchartComplex2016}, with a full set of topological invariants given in Ref.~\onlinecite{YaoTopological2017}. However, all of these works assume translational invariance, and none derives an explicit bulk-boundary correspondence. In the mathematics literature, Refs.~\onlinecite{Cedzichtopological2016,CedzichBulkedge2016} rigorously classified quantum walks in class~AIII without translational invariance, and arrived at a form of bulk-edge correspondence. However, the correspondence obtained therein is not able to distinguish between inherently dynamical edge modes and those which can arise in static systems, whose physical properties are often very different.

In the present work, we obtain a bulk-boundary correspondence which addresses many of the shortcomings of these previous studies. The new bulk invariant we introduce is physically motivated, locally computable, and applicable to systems both with and without disorder. The new edge invariants we define provide an explicit method for counting dynamical edge modes, and can be rigorously shown to equal the invariant of the bulk. In this way, our work forms an important milestone in the study and characterisation of FTI phases.

The structure of this paper is as follows. We begin, in Sec.~\ref{sec:Prelim}, by giving some background on time-dependent systems and the concepts we will use in obtaining the bulk-boundary correspondence. In Sec.~\ref{sec:bulk} we motivate and define a new bulk invariant we call `chiral flow', and give an example of its calculation for a model drive. We define two edge invariants in Sec.~\ref{sec:edge}, which we use to derive the bulk-edge correspondence in Sec.~\ref{sec:bulk-edge}. Finally, we summarise our results and conclude in Sec.~\ref{sec:conclusion}. To improve ease of reading, we use a stricter notion of locality than is necessary throughout the main text. Our results are extended to the more general case in the appendices, where we also provide further technical details on parts of the derivation.

\section{Preliminary Discussion\label{sec:Prelim}}
\subsection{Periodically Driven Systems}
We begin by reviewing some concepts from the study of time-dependent systems. A general time-periodic Hamiltonian satisfying $H(t+T)=H(t)$ generates the unitary time-evolution operator
\beq
U(t)&=&\mathcal{T}\exp\left[-i\int_0^tH(t')\dd t'\right],
\eeq
where $\mathcal{T}$ indicates time ordering (and we have set $\hbar=1$). In a time-independent system, we usually study the eigenstates of a static Hamiltonian in order to obtain information about the underlying topology. In a driven system, however, we are instead interested in the spectrum of the Floquet operator $U(T)$, the evolution operator after one complete period of driving. If the drive is topologically nontrivial, then this spectrum (in an open system) should exhibit protected boundary modes. These protected edge modes are driven-system analogues of the edge modes that arise (for example) in the quantum Hall effect or topological insulators.

Using an analogue of Bloch's theorem, the action of $U(T)$ may be written in terms of time-periodic eigenstates $\ket{\phi_n(t)}$ as
\beq
U(T) \ket{\phi_n(0)} = e^{-i\epsilon_n T}\ket{\phi_n(0)},
\eeq
where the quantities $\epsilon_n$ are known as quasienergies and are defined modulo $2\pi/T$. In many cases we can define an effective Floquet Hamiltonian $H_F$, which is related to the full-period time-evolution operator through
\beq
U(T)=\exp(-i H_F T).
\eeq
In order for $H_F$ to be well-defined, $U(T)$ must be gapped at some quasienergy $\epsilon_g$ so that a branch can be chosen when taking the logarithm. In addition, the branch cut must respect the underlying symmetries of the system. For topological phases, this feature generally restricts the utility of the Floquet Hamiltonian to closed systems, since open systems may have protected edge modes that lie in or across the quasienergy gap.

In Ref.~\onlinecite{RoyPeriodic2017} it was argued that the topology of a general periodic drive can have both dynamical and static components, and that these two components may be isolated from each other by a homotopic deformation of the unitary. To see this, we consider a closed-system unitary evolution whose end point is of the form $U(T)=\exp(-iH_FT)$, where $H_F$ is a static and local Hamiltonian. Then, we continuously deform the unitary evolution into a composition of a unitary loop $L$ (to be defined below) and a constant Hamiltonian evolution $C$, which is an evolution with the static Hamiltonian $H_F$. The dynamical component of the evolution is characterised by the loop $L$, a unitary evolution which (for a closed system) starts and ends at the identity,
\begin{equation}\label{eq1}
U(0)=U(T)=\id.
\end{equation}
In this way, the deformed evolution ($L$ followed by $C$) is a homotopic deformation of the original evolution.  This loop construction is outlined in more detail in Appendix~\ref{app:loop_from_unitary}. 

Overall, the bulk properties of an evolution can be described by studying the components $L$ and $C$ independently. The loop part of the evolution may be classified by a topological integer $n_L$, while the constant part of the evolution may be classified by a set of integers $n_{C_i}$, each associated with the $i$th gap in the constant Hamiltonian $H_F$ \cite{RoyPeriodic2017}.

The decomposition introduced above may be extended to an open system by removing terms from the generating Hamiltonian at each point in time that connect sites across a boundary cut. At the end of the evolution, these cuts may lead to nontrivial (protected) edge modes in the gaps in the quasienergy spectrum. Since the deformation is homotopic, any edge modes in the deformed evolution will be topologically equivalent to the edge modes in the original evolution. There is a one-to-one correspondence between the number of edge modes in each quasienergy gap (labelled $n_i$) and the integers characterising the bulk evolution, given by
\beq \label{loop-edge}
n_\pi&=&n_L\\
n_i&=&n_{C_i}+n_L,\nonumber
\eeq
where addition is taken modulo two if necessary \cite{RoyPeriodic2017}.\footnote{We note that for some symmetry classes, only the gaps at $\epsilon=0$ and $\epsilon=\pi$ are physically meaningful \cite{RoyPeriodic2017}.} In particular, $n_\pi$ counts the number of edge modes at $\epsilon=\pi$, which are inherently dynamical in nature and arise only loop component of the evolution is nontrivial. The remaining edge modes are similar to edge modes that arise in static Hamiltonians (although they may be affected by the loop component of the evolution).

A classification of evolutions by constant Hamiltonians is equivalent to the classification of static topological insulators and superconductors, and is well understood in the literature \cite{KitaevPeriodic2009,SchnyderClassification2008,KatsuraNoncommutative2016}. Since we are interested in Floquet topological phases, we will instead focus on inherently dynamical evolutions, described by unitary loops, and the associated dynamical edge modes at $\epsilon=\pi$. For this reason, we will assume throughout this paper that there is a bulk quasienergy gap at $\epsilon=\pi$, in which edge modes may appear in the open system. In the translationally invariant case, loop evolutions corresponding to all symmetry classes and dimensions were classified in Ref.~\onlinecite{RoyPeriodic2017}. In the present work, we study a certain class of loop evolutions which do not have translational symmetry.

\subsection{The Flow of a Unitary Matrix\label{sec:unitary_flow}}
In our study of dynamical phases, we will make use of a property of unitary matrices known as `flow', introduced by Kitaev in Ref.~\onlinecite{KitaevAnyons2006}. This property may be defined for any unitary operator, although we will later apply it to the special case of unitary loop evolutions. 

\begin{figure}[t]
\includegraphics[scale=0.65]{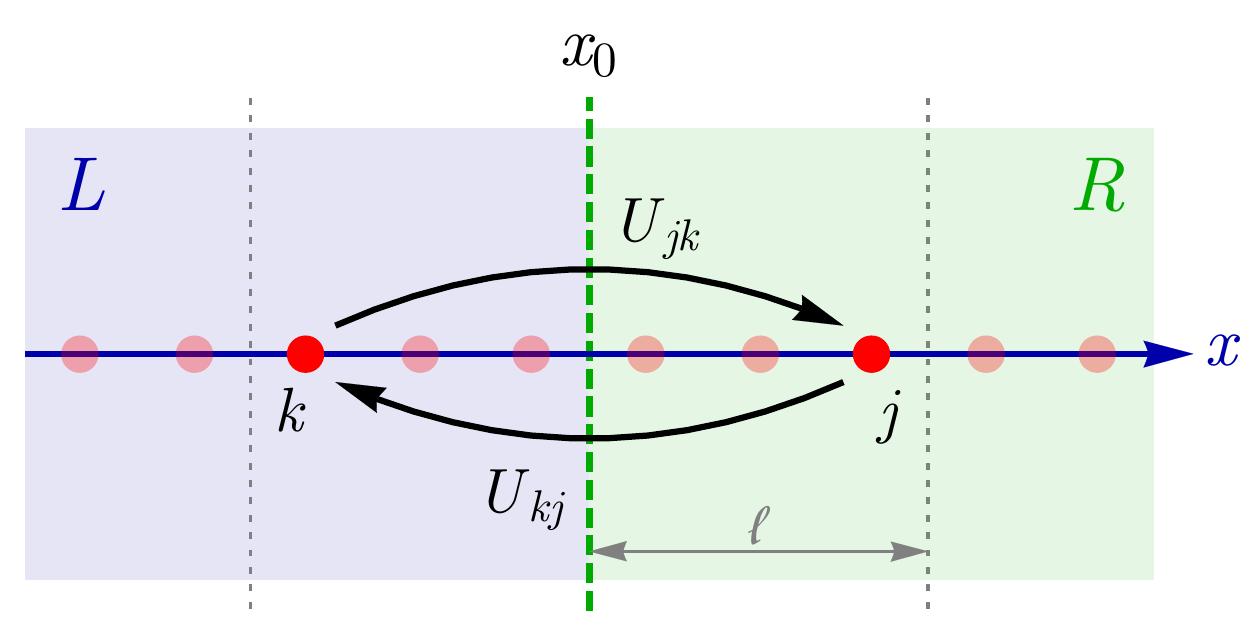}
\caption{The flow of a unitary matrix. Vertical dashed green line indicates the coordinate of the cross section $x_0$, defining regions $L$ and $R$. Red points indicate 1D lattice sites, while black arrows represent the current between sites $j$ and $k$ across the cut. The flow is defined by summing over all particle currents that cross the cut, as in Eq.~\eqref{eq:flow}. If the unitary is strictly local, only sites within a distance $\ell$ of the cut will contribute to the flow, delimited by grey dotted lines.\label{fig:gen_flow}}
\end{figure}

To define this quantity, we consider a noninteracting unitary matrix $U=(U_{jk})$, where $j$ and $k$ can be interpreted as labelling sites on a (formally infinite) one-dimensional lattice. Explicitly, the matrix elements $U_{jk}$ determine a unitary operator through the definition
\beq
\hat{U}&=&\sum_{jk}U_{jk}c^\dagger_{j}c^\ph_{k},
\eeq
where $c^\dagger_{j}$ ($c_j$) creates (annihilates) a single boson or fermion on site $j$. An intuitive notion of `current' from position $k$ to position $j$ induced by the unitary operator may then be defined as
\beq
f_{jk}=\abs{U_{jk}}^2-\abs{U_{kj}}^2,
\eeq
which is the difference in hopping probabilities between the two sites. In analogy with electric current, the one-dimensional flow of a unitary matrix is the total current through a cross section \cite{KitaevAnyons2006}, which may be written explicitly (for a cross section at coordinate $x_0$) as
\begin{equation}\label{eq:flow}
F(U)=\sum_{j\geq x_0}\sum_{k<x_0}f_{jk}.
\end{equation}
This is shown schematically in Fig.~\ref{fig:gen_flow}.

Following Ref.~\cite{KitaevAnyons2006}, we now introduce a projector $P_R$ for the half axis $x\geq x_0$ and a projector $P_L$ for the other half axis $x< x_0$, so that $P_L=\id-P_R$. In terms of these projectors, the flow of a unitary matrix may equivalently be rewritten as
\begin{equation}\label{flow_index}
\begin{split}
F(U)&=\textrm{Tr}(U ^\dag P_RUP_L-U ^\dag P_L UP_R)\\
&=\textrm{Tr}(U ^\dag P_RU-P_R)\\
&=\textrm{Tr}(U ^\dag \comm{P_R}{U}).
\end{split}
\end{equation}
It may be verified that these expressions reduce to Eq.~\eqref{eq:flow}. Importantly, in these expressions we cannot use the cyclic property of the trace $\textrm{Tr}(AB)= \textrm{Tr}(BA)$, since this holds only if one of the matrices has a finite number of nonzero elements \cite{KitaevAnyons2006}. 

In a physical system, the underlying Hamiltonian must satisfy certain locality constraints, which in turn places constraints on the form of the time-evolution operator. We can use these properties to our advantage when calculating the flow of a unitary operator that corresponds to time evolution. The usual definition of a local operator is one whose matrix elements decay exponentially (or faster) with the distance between the sites involved (see, for example, Ref.~\onlinecite{GrafBulk2018}). To simplify our discussion in the main text, however, we will assume that the time-evolution operator is \emph{strictly} local, i.e. that $U_{jk}=0$ for $\abs{j-k}>\ell$ beyond some localisation length $\ell$ (where $j$ and $k$ label unit cell positions). In Appendix~\ref{app:exponentially_decaying} we extend our arguments to the more general definition of locality.

Under this assumption of strict locality, it is clear that only the regions close to the cross section will contribute to the calculation of flow in Eq.~\eqref{flow_index}. Setting the cross section coordinate to be $x_0=0$, we can therefore restrict our projectors to the relevant interval $[-\ell,\ell]$. Specifically, we define the projector $P_L^\ell$ for the region $[-\ell,0)$ and the projector $P_R^\ell$ for the region $[0,\ell)$ and substitute $P_L\to P_L^\ell$ and $P_R\to P_R^\ell$ in Eq.~\eqref{flow_index}. It is clear that the result will not be affected by this truncation, and we arrive at the expression
\begin{equation}\label{eq:flow_ell}
F(U)=\textrm{Tr}(U^\dag P_R^\ell U P_L^\ell-U^\dag P_L^\ell U P_R^\ell).
\end{equation}
With this truncation, we have an expression for $F(U)$ that can now be applied to finite systems.

To develop some intuition for this index, we now consider two simple examples. First, we take the unitary operator to be the identity, $U=\mathbb{I}$. It is clear that the flow will be zero in this case, since each projector commutes with $U$ in Eq.~\eqref{eq:flow_ell} and we can apply the result $P_L^\ell P_R^\ell=0$. This is reflective of the fact that that unitary operator $\id$ does not involve any particle current between the two sides of the cut.

As a second example we consider the unitary $U=\hat{t}$, where $\hat{t}$ is the shift operator having action
\begin{equation}\label{eq:shift}
\hat{t}\ket{j}=\ket{j+1},
\end{equation}
with $\ket{j}$ a state localised on site $x=j$. For this operator the localisation length is $\ell=1$. Writing $P_L^\ell=\ket{-1}\!\bra{-1}$ and $P_R^\ell=\ket{0}\!\bra{0}$, we see that
\beq
F(U)&=&\textrm{Tr}\left[\left(\hat{t}^\dag \ket{0}\!\bra{0} \hat{t} \ket{-1}\!\bra{-1}\right)-\left(\hat{t}^\dag \ket{-1}\!\bra{-1}\hat{t} \ket{0}\!\bra{0}\right)\right]\nonumber\\
&=&1.
\eeq
This quantifies the non-zero current of particles across the cut effected by the operator $\hat{t}$.

The flow of a unitary operator, using any of the definitions above, is quantized to take integer values \cite{KitaevAnyons2006}. It is therefore robust to any continuous change of the system, and can be used for the purposes of topological classification. When $U$ possesses translational symmetry, we may rewrite the expression for flow using the Fourier transform,
\begin{equation} \label{eq:Fourier_def}
\widehat{U}(q)=\sum_x U_{0x} e^{iqx},
\end{equation}
finding
\begin{equation}\label{eq:winding}
F(U)=\textrm{tr}\left[U^\dagger\comm{X}{U}\right]=\frac{i}{2\pi}\int_{-\pi}^{\pi}\dd q\,\textrm{Tr}\left[\widehat{U}(q)^\dag \frac{d\widehat{U}}{dq}\right].
\end{equation}
In this expression, we distinguish between `tr', meaning the trace per unit cell, and `Tr', which is the usual matrix trace \cite{KitaevAnyons2006}. In this way, we see that for translationally invariant systems, the flow is equal to the familiar momentum-space winding number $w[U]$ that captures the topology of the mapping from the space $S^1$ to the space of unitary matrices.

In this paper, we aim to find and classify time-evolution operators with nontrivial unitary flow. However, many of the systems we might hope would host such phases can be shown to have a trivial flow index. In particular, a 1D unitary generated by a local 1D Hamiltonian can be shown to always have zero flow index \cite{GrossIndex2012}. In addition, for a unitary operator with a finite number of nonzero elements, we can use the cyclic property of the trace to show that Eq.~\eqref{flow_index} vanishes. 

One way to achieve a nonzero flow is to consider higher-dimensional systems, where, for example, robust chiral edge modes may exist at the boundary of a 2D periodic drive \cite{RudnerAnomalous2013}. In this work, we will instead consider inherently 1D systems with a protected chiral symmetry, which leads to a definition of `chiral flow'.

\section{Driven Systems with Chiral Symmetry\label{sec:bulk}}
\subsection{Chiral Symmetry}
In this section, we will study topological drives with chiral symmetry, corresponding to Class~AIII of the AZ classification scheme \cite{HeinznerSymmetry2005,AltlandNonstandard1997,RoyPeriodic2017}. We will introduce a notion of chiral flow that can be used to characterise such systems and describe a nontrivial model drive that may be used to generate phases with different chiral flow indices. 

We begin by recalling that a system has chiral symmetry if its (time-dependent) Hamiltonian satisfies the relation
\beq
CH(t)C^{-1}=-H(-t)\label{eq:ham_chiral}
\eeq
for some chiral symmetry operator $C$, which is a unitary operator satisfying $C^2=\id$ \cite{RoyPeriodic2017}. The eigenvalues of such an operator are $\pm1$, and so we can write it in diagonal form as
\beq
C&=&\left(\begin{array}{cc}
\id & 0\\
0 & -\id
\end{array}\right)=C_+-C_-,\label{eq:c_diagonal}
\eeq
where $C_\pm$ are projectors onto the $\pm1$ eigenspace. In this basis, the instantaneous Hamiltonian is off-diagonal. It follows from Eq.~\eqref{eq:ham_chiral} that chiral symmetry acts on the unitary time-evolution operator as
\beq
CU(t)C^{-1}&=&U(T-t)U^\dagger (T),\label{eq:chiral_unitary}
\eeq
where we have used the fact that the Hamiltonian is periodic in time with period $T$ \cite{RoyPeriodic2017}. At the end of one cycle, the time-evolution operator satisfies
\beq
CU(T)C^{-1}&=&U^\dagger (T).\label{chiral_unitary_T}
\eeq

\subsection{A Model Drive with Chiral Symmetry\label{sec:model_chiral_drive}} 
We now build a model drive with chiral symmetry that has nontrivial flow properties. Inspired by the static Su-Schrieffer-Heeger (SSH) model \cite{SuSolitons1979}, we take a 1D bipartite chain of $N$ unit cells, with sublattices labelled `A' and `B'. For a closed chain, we define the SSH-like Hamiltonian
\beq
H_{\rm SSH}&=&v\sum_{m=1}^N \bigg[\ket{m,B}\bra{m,A}+\hc\bigg] \label{eq:SSH_Ham}\\
&+&w\sum_{m=1}^{N} \bigg[\ket{(m+1) N,A}\bra{m,B}+\hc\bigg], \nonumber
\eeq
where $\ket{m,\alpha}$ denotes a state of the chain where the particle is on sublattice $\alpha$ in unit cell $m$. The parameter $v$ controls the hopping of particles between sublattice A and B within the same cell, while the parameter $w$ controls the hopping between unit cells. In this static case, the Hamiltonian is trivial (topological) when $|v|>|w|$ ($|w|>|v|$) \cite{SuSolitons1979}. It may be verified that this Hamiltonian satisfies the chiral symmetry constraint
\be
CH_{\rm SSH}C^{-1}&=&-H_{\rm SSH},
\ee
where $C$ is defined in the canonical way through $C=\prod_j\tau^z_j$, and where $\tau_j^z$ is a Pauli $z$-matrix acting in the sublattice space on site $j$.

\begin{figure}[t]
\includegraphics[scale=0.5]{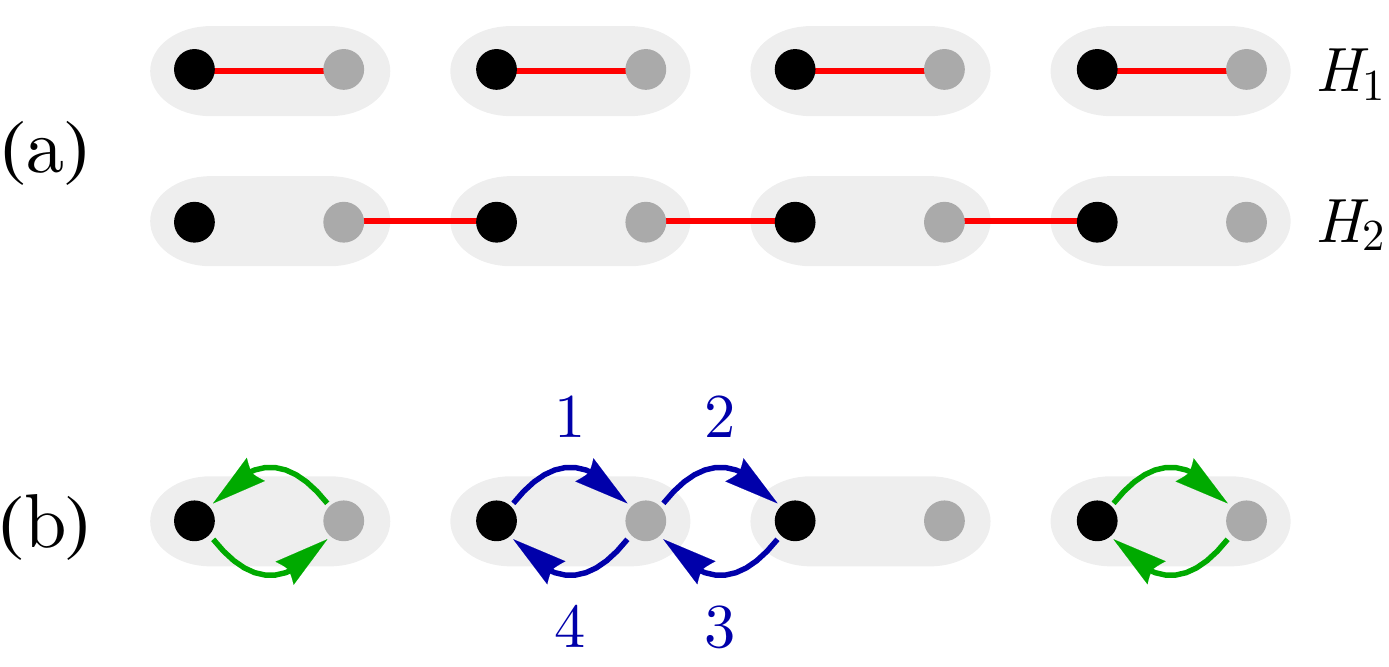}
\caption{Schematic picture of the model drive given in Eqs.~\eqref{eq:composition_TRS}~and~\eqref{eq:model_h}. (a) Driving protocol. The phase with Floquet edge states is obtained by driving with a trivial Hamiltonian $H_1$ and a nontrivial Hamiltonian $H_2$ in turn. Black (gray) points are sites on sublattice A (B). For each Hamiltonian, hopping occurs between sites indicated by red lines. Gray background indicates unit cells. (b) Drive action. Particles in the bulk move in closed trajectories indicated by the blue arrows, while particles on the boundary follow the green arrows and gain a phase of $\pi$. \label{fig:model_drive}}
\end{figure}

For our model drive, we take a piecewise constant Hamiltonian of the form
\beq
H(t)&=&\left\{\renewcommand\arraystretch{1.2}\begin{array}{ccc}
H_1 && 0\leq t<\frac{1}{4} T\\
H_2 && \frac{1}{4} T \leq t< \frac{1}{2} T\\
H_2 && \frac{1}{2} T \leq t< \frac{3}{4} T\\
H_1 && \frac{3}{4} T \leq t<  T,
\end{array}\right.,\label{eq:composition_TRS}
\eeq
where
\beq
H_1&=&\frac{2\pi}{T}\sum_{m=1}^N (\ket{m,B}\bra{m,A}+h.c.) \label{eq:model_h}\\
H_2&=&-\frac{2\pi}{T}\sum_{m=1}^{N-1} (\ket{m+1 ,A}\bra{m,B}+h.c.) \qquad  \nonumber
\eeq
are of the form in Eq.~\eqref{eq:SSH_Ham}.\footnote{The relative minus sign is added for convenience so that hopping phases gained during steps one and two cancel out in the bulk.} The hopping terms of the drive are indicated in Fig.~\ref{fig:model_drive}(a). 

In each of the four steps of the drive, a particle moves with probability one between neighboring sites, following the trajectory shown in Fig.~\ref{fig:model_drive}(b). After a complete cycle, each particle in the closed system returns to its initial position, so that $U(T)=\id$ and the unitary is a loop. After cutting the chain open, however, some terms from the Hamiltonian are removed, and one particle on each edge no longer moves during the second and third steps. At the end of the evolution, these particles have instead gained a phase of $\pi$, and will show up in the quasienergy spectrum as protected edge modes at $\epsilon=\pi$.

\subsection{Winding Number Invariant}
\begin{figure}[t]
\includegraphics[scale=0.55]{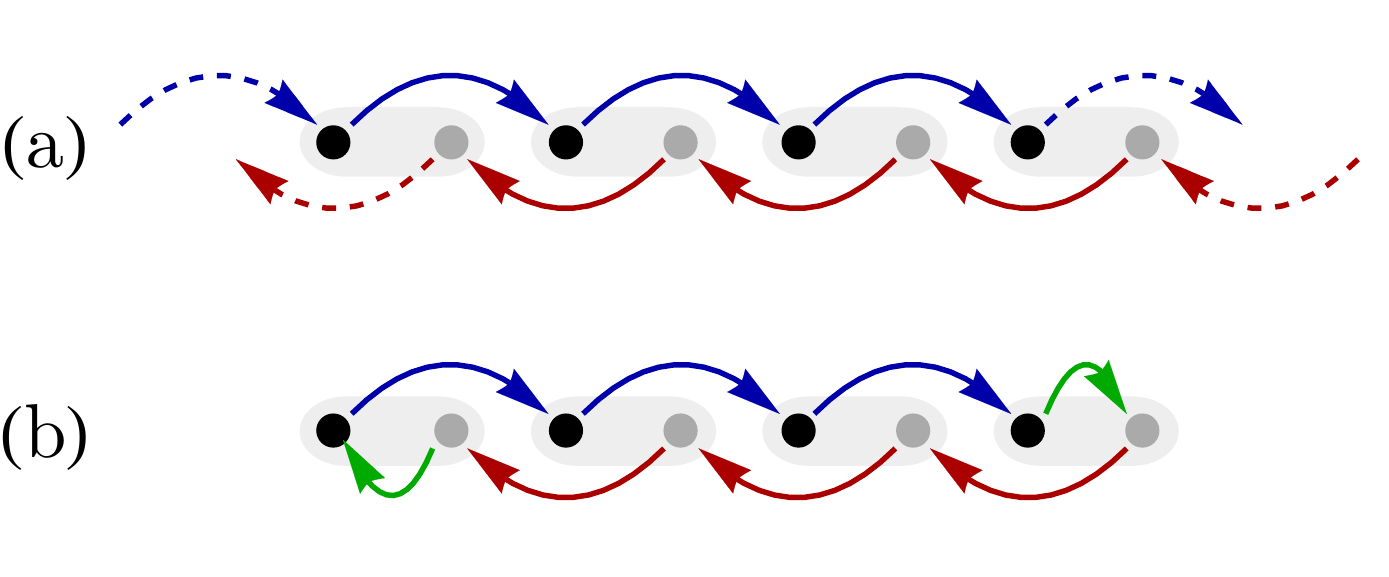}
\caption{Schematic picture of the chiral flow for the model drive introduced in Sec.~\ref{sec:model_chiral_drive}. (a) The chiral flow in the closed system generated by $U(T/2)$. Black (grey) circles indicate sublattice A (B). Blue (red) lines show the motion of particles on sublattice A (B) after a half period. (b) The chiral flow in the corresponding open system after a half period. Note that on each edge, particles are forced to hop between sublattices (green arrows). \label{fig:chiral_flow}}
\end{figure}

We now obtain a classification of this model drive which may be extended to more general systems with chiral symmetry. We first observe that at $t=T/2$ in the model drive, particles that started on an A site have moved to another A site (unless they are near the boundary) and similarly, particles that started on sublattice B have moved to another B site. More specifically, particles appear to `flow' within each sublattice, and the action of the half-period evolution $U(T/2)$ resembles a combination of shift operators $\hat{t}\oplus \hat{t}^{\dagger}$, introduced in Eq.~\eqref{eq:shift}. This is shown schematically in Fig.~\ref{fig:chiral_flow}(a).

This sublattice decoupling is a general feature of Floquet drives with chiral symmetry at the half-period point. We can see this by substituting $t=T/2$ into Eq.~\eqref{eq:chiral_unitary} to find
\beq\label{eq:half_period_full_period_relation} \label{U(T) and U(T/2)}
U^\dagger\left(T/2\right)CU\left(T/2\right)C^{-1}&=&U^\dagger\left(T\right).
\eeq
If the drive is a loop, then
\begin{equation}\label{eq19}
CU\left(T/2\right)C^{-1}=U\left(T/2\right),
\end{equation}
which means that $U(T/2)$ commutes with the chirality operator and hence is block-diagonal in the chiral basis,
\begin{equation}\label{eq:block_diagonal}
U\left(T/2\right)=
\begin{pmatrix}
U_+&0\\0&U_-
\end{pmatrix}.
\end{equation}
For the model drive, we see that $U_+=\hat{t}$ and $U_-=\hat{t}^{\dagger}$.

Now, we know that any translationally symmetric (closed-system) unitary in 1D has a winding number as given in Eq.~\eqref{eq:winding}. Applying this formula to the block-diagonal unitary $U(T/2)$, we find
\beq\label{eq24}
w[U(T/2)]&=&w[U_+]+w[U_-]=0,
\eeq
which must vanish because the evolution is one-dimensional: Specifically, the winding number is a homotopy invariant and $U(t)$ is smooth, and so $w[U(t)]$ must be independent of time \cite{FruchartComplex2016}. Then, since $w[U(0)]=w[\id]=0$ at the beginning of the evolution, it follows that $w[U(T/2)]=0$ too.

However, we observed for the model drive in Sec.~\ref{sec:model_chiral_drive} a `chiral flow' within each sublattice for which $w\neq0$, and so in general $w[U_+]$ and $w[U_-]$ might not be zero individually. We identify the quantity $w[U_+]$ as the topological invariant $\nu[U]$ for a chiral Floquet system, which may be written in full as
\beq\label{eq25}
\nu[U]&=&\frac{i}{2\pi} \int_{-\pi}^{\pi} \dd k\, \textrm{tr}\bigg[U_+^{-1}(k)\partial_k U_+(k)\bigg]\\
&=&\frac{i}{4\pi} \int_{-\pi}^{\pi} \dd k\, \textrm{tr}\bigg[CU^{-1}(k,T/2)\partial_k U(k,T/2)\bigg].\nonumber
\eeq
In the second line we have inserted the chiral symmetry operator and used the fact that $w[U_+]=-w[U_-]$. For the model drive above, it may be verified that $\nu[U]=1$. Other integer values of $\nu[U]$ can be obtained by running this model drive in sequence (being sure to preserve chiral symmetry), or by running the model drive in reverse. 

Since the winding number is quantised to take integer values, it is robust to local perturbations and is a well-defined topological index for chiral symmetric Floquet systems with translational invariance. The topological invariant for 1D Floquet systems in class~AIII has previously been expressed in this form in Ref.~\onlinecite{FruchartComplex2016}.

\subsection{Chiral Flow Invariant\label{sec:real_space_invariant}}
The winding number above functions as a topological invariant for Floquet systems with both chiral symmetry and translational invariance. We now seek to construct a real-space topological invariant that continues to hold even in disordered systems, using the connection between flow and winding number discussed in Sec.~\ref{sec:unitary_flow}.

From Eq.~\eqref{eq:winding}, we see that the winding number $w(U_+$) is related to the flow of the unitary $U_+$, which acts only on the positive chirality eigenspace. We can therefore use Eqs.~\eqref{flow_index}~and~\eqref{eq:flow_ell} to replace $w[U_+]$ with $F[U_+]$, so that
\beq \label{bulk_invariant}
\nu[U]&=&\textrm{\textrm{Tr}}(U_+^{-1} P_RU_+P_L)-\textrm{\textrm{Tr}}(U_+^{-1} P_LU_+P_R)\nonumber\\
&=&\textrm{\textrm{Tr}}(U_+^{-1} P_R^\ell U_+P_L^\ell)-\textrm{Tr}(U_+^{-1} P_L^\ell U_+P_R^\ell).
\eeq
This describes the flow from the left side of a cut to the right, using the definitions of projectors introduced in Sec.~\ref{sec:unitary_flow}. In the second line, we have again assumed that the unitary operator is \emph{strictly} local, so that the truncation to a region of width $2\ell$ around the cut has no effect on the calculation of $\nu[U]$. In Appendix~\ref{app:exponentially_decaying}, we consider the more general case where $U$ is only exponentially localised.


As in the translationally invariant case, the total flow of the unitary must be zero and so $F[U_+]=-F[U_-]$. We can therefore write the real-space invariant more symmetrically as
\beq
\nu[U]&=&\frac{1}{2}\bigg[\textrm{\textrm{Tr}}\left(CU^{-1} P_RUP_L\right)-\textrm{\textrm{Tr}}\left(CU^{-1} P_LUP_R\right)\bigg]\nonumber\\
&=&\frac{1}{2}\bigg[\textrm{\textrm{Tr}}\left(CU^{-1} \comm{P_R}{U}\right)\bigg]\label{eq:chiral_flow}\\
&=&\frac{1}{2}\bigg[\textrm{\textrm{Tr}}\left(CU^{-1} P_R^\ell UP_L^\ell \right)-\textrm{\textrm{Tr}}\left(CU^{-1} P_L^\ell UP_R^\ell \right)\bigg] \nonumber
\eeq
where we have used the shorthand $U=U(T/2)$.

These (equivalent) expressions for $\nu[U]$ define what we refer to as \emph{chiral flow}, a bulk topological invariant that quantifies the particle flow on a single sublattice at the midpoint of a Floquet evolution belonging to class~AIII. For the model drive introduced in Sec.~\ref{sec:model_chiral_drive}, this flow is evident from the form of the unitary operator at $t=T/2$, but $\nu[U]$ may be calculated for any evolution in the symmetry class. In particular, the real-space expression for the chiral flow is applicable to unitary evolutions with disorder.

\begin{figure}[t]
\includegraphics[scale=0.5]{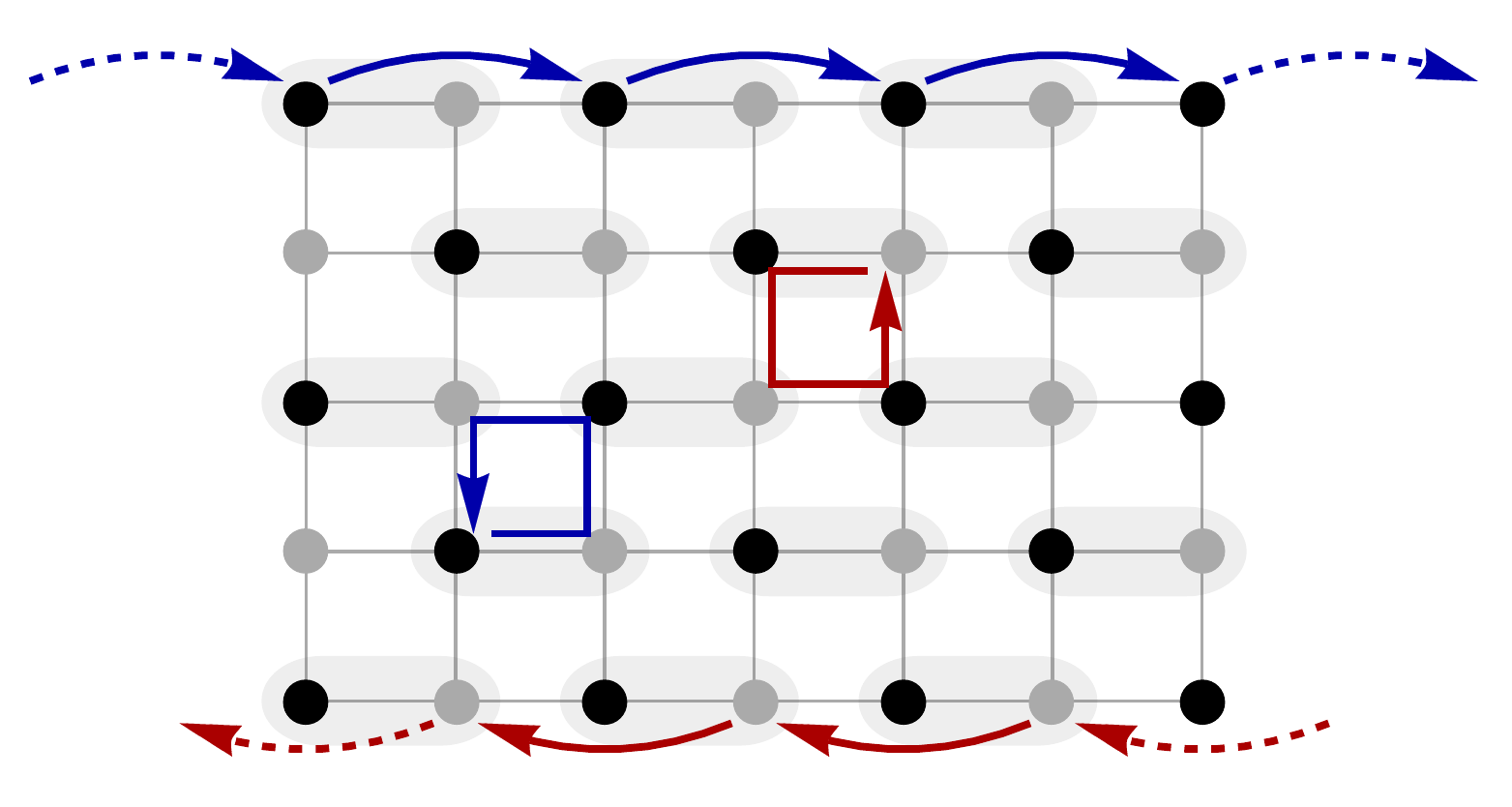}
\caption{Action of the 2D class~A drive introduced in Ref.~\onlinecite{RudnerAnomalous2013}. During each step, particles hop between sites on a bipartite lattice following the paths indicated by red and blue arrows. Sublattices are indicated by gray and black points, while gray shaded regions indicate unit cells. After a complete cycle, a particle in the bulk returns to its initial position, while a particle at the edge is translated by one unit cell. The 1D class~AIII model we introduce in Sec.~\ref{sec:model_chiral_drive} can be obtained by collapsing this 2D model onto a single chain. \label{fig:rudner}}
\end{figure}

Before concluding this section, we note that there is another way of interpreting the chiral flow in this model drive. In Ref.~\onlinecite{RudnerAnomalous2013}, the authors construct a 2D Floquet loop drive belonging to class~A (which has no protecting symmetries). Under the action of their drive, a particle in the bulk will follow a closed path around a square plaquette, returning to its initial position. However, a particle at the boundary will be unable to complete a closed path, and will instead propagate along the edge, as shown in Fig.~\ref{fig:rudner}.

At $t=T/2$, our 1D class~AIII model drive exhibits bulk chiral flow that looks very similar to the edge behaviour (at $t=T$) of the 2D class~A model of Ref.~\onlinecite{RudnerAnomalous2013}. In fact, our class~AIII model can be interpreted as an open-system class~A drive collapsed down to a single layer. In this interpretation, the edge modes of the model of Ref.~\onlinecite{RudnerAnomalous2013} undergo chiral flow at the boundary.

\section{Dynamical Edge Modes in Driven Systems with Chiral Symmetry\label{sec:edge}}
\subsection{Protected Dynamical Edge Modes in the Model Drive}
In the previous section, we introduced chiral flow as a robust topological quantity that describes the bulk properties of a Floquet evolution in class~AIII halfway through the driving cycle. However, topological phases also exhibit robust edge behaviour that is closely related to the physics in the bulk, a feature known as bulk-boundary correspondence. In this section, we study the form of the dynamical edge modes present at the end of a topological drive in class~AIII, and introduce an invariant that may be used to count them. This will be used in Sec.~\ref{sec:bulk-edge} when we derive an explicit bulk-edge correspondence.

In general, protected edge states arise at interfaces between topological phases where bulk topological numbers change. In this paper, we will mostly consider edge modes at the boundary of an open system, which may be viewed as an interface with the (topologically trivial) vacuum. As motivated in Sec.~\ref{sec:Prelim}, inherently dynamical edge modes are associated with bulk unitary loops, and occur at quasienergy $\epsilon=\pi$.

As an example, we first revisit the model drive introduced in Sec.~\ref{sec:model_chiral_drive}. We recall that halfway through the drive, the time-evolution operator takes the block-diagonal form
\be
U_c(T/2)&=&\left(\begin{array}{cc}
\hat{t} & 0\\
0 & \hat{t}^\dagger
\end{array}\right),
\ee
where $\hat{t}$ ($\hat{t}^\dagger$) is the unit translation operator to the right (left), and each translation operator acts within a single sublattice. We have added the subscript $c$ to emphasise that the unitary operator here is for the closed system. In this way, the unitary $U_c(T/2)$ generates a chiral flow with index $\nu[U]=1$, shown schematically in Fig.~\ref{fig:chiral_flow}.

In order to find the number and form of the edge modes of the model, we must evolve with the drive in an open system until $t=T$. Using Eqs.~\eqref{eq:SSH_Ham}~and~\eqref{eq:composition_TRS}, we find
\beq
U_o(T)&=&-\ket{1,B}\bra{1,B}-\ket{N,A}\bra{N,A} \label{eq:model_drive_edge}\\
&&+\sum_{m=1}^{N-1} \ket{m,A}\bra{m,A}+\sum_{m=2}^{N} \ket{m,B}\bra{m,B}, \nonumber
\eeq
where the subscript $o$ indicates this is the evolution for the open system (i.e. with terms in the generating Hamitonian which connect sites across the boundary omitted). Then, writing
\beq
U_o(T)&=&\sum_ne^{-i\epsilon_nT}\ket{\phi_n}\!\bra{\phi_n},
\eeq
we see that there is a single edge mode at $\epsilon=\pi$ on each boundary. Focussing on the right-hand edge at $x=N$, we find that there is one edge mode with wavefunction $\ket{N,A}$, and that the total number of $\pi$ modes is $n_\pi=1$. In general, we write the net number of $\pi$ edge modes at a single boundary as
\beq
n_\pi&=&n_\pi^A-n_\pi^B,\label{eq:n_edge_modes}
\eeq
which is the number of edge modes on sublattice A minus the number of edge modes on sublattice B. This definition is justified in that a pair of degenerate states at the same edge on different sublattices can be gapped out by a chiral-symmetric Hamiltonian acting only at the edge, as proved below in Sec.~\ref{sec:edge_invariants}.

For our model drive, it follows that at the right edge $n_\pi^R=1$, while at the left edge $n_\pi^L=-1$. In this way, at least for the model drive, we see that the number of edge modes is equal to the \emph{change} in the bulk chiral flow invariant across the interface,
\beq
n_\pi=n_\pi^A-n_\pi^B=\Delta\nu.
\eeq
We will show below that this bulk-edge correspondence holds in general.

\subsection{Edge Invariants\label{sec:edge_invariants}}
\begin{figure}[t]
\includegraphics[scale=0.4]{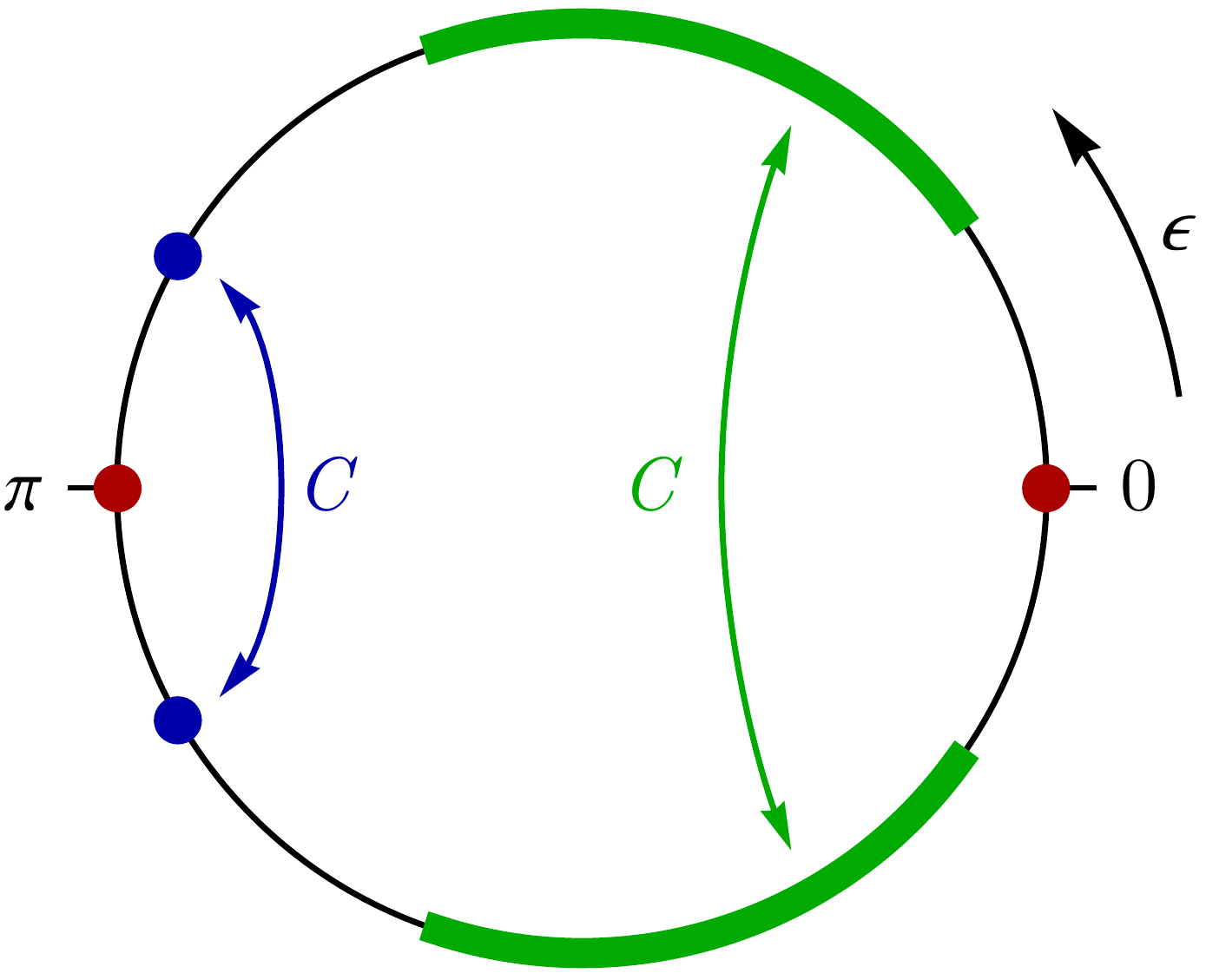}
\caption{Schematic quasienergy spectrum for a Floquet operator $U(T)$ with chiral symmetry. Since quasienergies are defined modulo $2\pi$, the spectrum may be visualised on the unit circle. The chiral symmetry operator $C$ maps states with quasienergy $\epsilon$ onto states with quasienergy $-\epsilon$ (green band and blue points). In this way, the spectrum is symmetric about $\epsilon=0$ and $\epsilon=\pi$. States at these special values map onto states with the same quasienergy under the action of $C$, and possibly map onto themselves (red points). \label{fig:quasienergy}}
\end{figure}

We now extend this discussion of edge modes to more general drives with chiral symmetry. First, we note from Eq.~\eqref{chiral_unitary_T} that the quasienergy spectrum of a chiral-symmetric Floquet operator $U(T)$ is symmetric about $\epsilon=0$ and $\epsilon=\pi$. Specifically, if $\ket{\phi_n}$ is an eigenstate of $U(T)$ with quasienergy $\epsilon_n$~(mod~$2\pi$), then using Eq.~\eqref{chiral_unitary_T} we can write
\beq
U(T)C\ket{\phi_n}=CU^\dagger(T)\ket{\phi_n}=e^{-i\epsilon_n}C\ket{\phi_n},
\eeq
which shows that $C\ket{\phi_n}$ is also an eigenstate of $U(T)$ with quasienergy $-\epsilon_n$~(mod~$2\pi$). In this way, eigenstates at $\epsilon=0$ and $\epsilon=\pi$ are special, in that the chiral symmetry operator maps them onto states with the same quasienergy. These spectral properties are illustrated in Fig.~\ref{fig:quasienergy}.

For the open-system Floquet operator $U_o(T)$ of the model drive given in Eq.~\eqref{eq:model_drive_edge}, we see that each edge state has support on a single sublattice, and is mapped onto itself under the action of $C$. In fact, eigenstates at $\epsilon=\pi$ can always be chosen to have support on a single sublattice, as we now show. First, we write the projector onto sublattice A/B as
\beq
P_{A/B}&=&\frac{\id\pm C}{2},\label{eq:pab_chiral}
\eeq
which follows from Eq.~\eqref{eq:c_diagonal}. Then, given an eigenstate
\beq
U_o(T)\ket{\phi}&=&-\ket{\phi},
\eeq
we see that
\beq
U_o(T)P_A\ket{\phi}&=&U_o(T)\left[\frac{\id+C}{2}\right]\ket{\phi}\nonumber\\
&=&\frac{1}{2}\left[U_o(T)+CU_o^\dagger(T)\right]\ket{\phi}\nonumber\\
&=&-P_A\ket{\phi},
\eeq
where we have made use of Eq.~\eqref{chiral_unitary_T}. In this way, $P_A\ket{\phi}$ is either an eigenstate of $U_o(T)$ at $\epsilon=\pi$ or $\ket{\phi}$ is annihilated by $P_A$. In either case, we can split the state $\ket{\phi}$ into components $P_A\ket{\phi}$ and $P_B\ket{\phi}$, each of which has support on a single sublattice or vanishes. A state with support on a single sublattice is mapped onto itself by the action of $C$.

Now, a dynamical topological phase is indicated by the presence of \emph{protected} edge modes at $\epsilon=\pi$. In order to be protected, it should not be possible to gap out the edge modes with a local, symmetry-respecting evolution acting only in the edge region. We now demonstrate that for edge modes to be protected, they must all have support on the same sublattice. Specifically, we will show that a pair of edge modes (at the same edge) with support on different sublattices may be gapped out, providing justification for the edge-mode counting defined in Eq.~\eqref{eq:n_edge_modes}.

We assume we have two eigenstates $\ket{\phi_A}$ and $\ket{\phi_B}$ at quasienergy $\epsilon=\pi$, with support on sublattice A and B, respectively. While the states do not need to have support on the same sites, they should each be localised to the same boundary. We then consider the local, chiral-symmetric Hamiltonian
\beq
H'&=&\ket{\phi_A}\!\bra{\phi_B}+\ket{\phi_B}\!\bra{\phi_A},
\eeq
which generates the evolution,
\beq
e^{-itH'}&=&\left(\begin{array}{cc}
\cos(t) & -i\sin(t)\\
-i\sin(t) & \cos(t) 
\end{array}\right),
\eeq
where we have used the basis $\{\ket{\phi_A},\ket{\phi_B}\}$. To form a chiral-symmetric unitary evolution, we prepend and append this new evolution to the original unitary $U_o(T)$. Considering the action of this new evolution on the edge-state subspace, we find
\beq
e^{-itH'}U_o(T)e^{-itH'}&=&e^{-itH'}\left(\begin{array}{cc}
-1 & 0\\
0 & -1
\end{array}\right)e^{-itH'}\nonumber\\
&=&\left(\begin{array}{cc}
-\cos(2t) & i\sin(2t)\\
i\sin(2t) & -\cos(2t)
\end{array}\right).
\eeq
This new unitary evolution has quasienergies $\epsilon=\pi\pm2t$, and so even for an infinitesimal perturbation, the edge states are mixed and gap out. In this way, a pair of edge modes at $\epsilon=\pi$ on different sublattices can be gapped out by a local, symmetric perturbation, and are not protected. The number of protected edge modes at a given edge is the \emph{difference} between the number of edge modes on sublattice A and the number of edge modes on sublattice B, as defined in Eq.~\eqref{eq:n_edge_modes}.

With these definitions, we can now obtain a general expression for the number of protected edge modes present at the boundary of an arbitrary chiral drive. An open-system drive (derived from a unitary loop evolution) will have a thermodynamically large number of eigenstates at $\epsilon=0$ corresponding to states in the bulk, and a smaller number of states with $\epsilon\neq0$ near each boundary. The number of edge modes is the net number of eigenstates on a single sublattice at quasienergy $\epsilon=\pi$ on a single boundary.

We project to just the right-hand boundary with the real-space projector $P_R$, where the boundary region need not be exact but should include all states on the right-hand edge with $\epsilon\neq0$. We also define a projector $P_\pi$ onto the space of states with quasienergy $\epsilon=\pi$ (we give an explicit expression for this operator below). In terms of these projectors, the number of protected dynamical edge modes at the right-hand edge is given by
\begin{equation}
n_\pi^R=n_\pi^{R,A}-n_\pi^{R,B}=\mathrm{Tr}\left[P_AP_\pi P_R\right]-\mathrm{Tr}\left[P_BP_\pi P_R\right].
\end{equation}
Recalling that the chiral symmetry operator takes the form $C=P_A=P_B$, this expression can be rewritten as
\beq
n_\pi^R[U]=\mathrm{Tr}\left[CP_\pi P_R\right]=-\frac{1}{2}\mathrm{Tr}\left[C\left(U_o(T)-\id\right) P_R\right],\label{eq:edge_invariant}
\eeq
where in the final equality we have replaced $P_\pi=-\frac{1}{2}\left[U_o(T)-\id\right]$ under the trace. 

This replacement can be justified as follows, by expanding $U_o(T)$ in its basis of eigenstates,
\beq
U_o(T)&=&\sum_n e^{-i\epsilon_n}\ket{\phi_n}\!\bra{\phi_n}.
\eeq
First, subtracting the identity removes all states with $\epsilon=0$ from the expansion of $U_o(T)$, meaning that these states do not contribute to the trace. The states with $\epsilon=\pi$, however, have a coefficient of $-1$ in the expansion of $U_o(T)$, and so end up with a coefficient of $-2$ after subtracting the identity. States at $\epsilon=\pi$ will therefore each contribute $-2$ to the trace (if they are not annihilated by $P_R$ or $C$). Finally, if there are any states with $\epsilon\neq0$ and $\epsilon\neq\pi$, they must occur as chiral pairs with quasienergy $\pm\epsilon$. However, the chiral symmetry operator $C$ acts as $\sigma_x$ on these eigenvectors (mapping each state onto its chiral partner), and is therefore traceless in this subspace. 

Overall, the only subspace that contributes to the trace of $C(U_o(T)-\id)P_R$ is the $\pi$ eigenspace of the right-hand edge, and we divide by $-2$ to calculate the number of states in this subspace. As long as the chosen region $R$ is larger than the localisation length of  any edge modes, the trace will be integer valued. It follows that Eq.~\eqref{eq:edge_invariant} is a robust topological edge invariant which may be used to calculate the number of protected edge modes at the right-hand edge of any chiral-symmetric Floquet operator $U_o(T)$.

\section{Bulk-Edge Correspondence\label{sec:bulk-edge}}
\subsection{Bulk-Edge Correspondence at $t=T/2$}
In this section, we will prove that the bulk chiral flow invariant of a chiral unitary loop drive ($\nu[U]$) is equal to the number of protected edge modes at the right-hand edge at the end of the evolution ($n_\pi^{R}$). Our argument has two parts: first, we will show that a nonzero chiral flow in the bulk at $t=T/2$ leads to chiral-symmetry-breaking flow at the boundary, also at $t=T/2$. Then, we will show that this symmetry-breaking boundary flow at $t=T/2$ is responsible for nontrivial edge modes at the end of the cycle.

As in the previous section, it will be useful to distinguish between the closed-system evolution and the open-system evolution, which we write as $U_c(t)$ and $U_o(t)$, respectively. These two evolutions are identical apart from in a finite (Lieb-Robinson bounded) region near the edges. In particular, since we are considering unitary loop evolutions, $U_c(T)=\id$ everywhere, while $U_o(T)$ is the identity away from the boundary regions. 

\begin{figure}[t]
\includegraphics[scale=0.55]{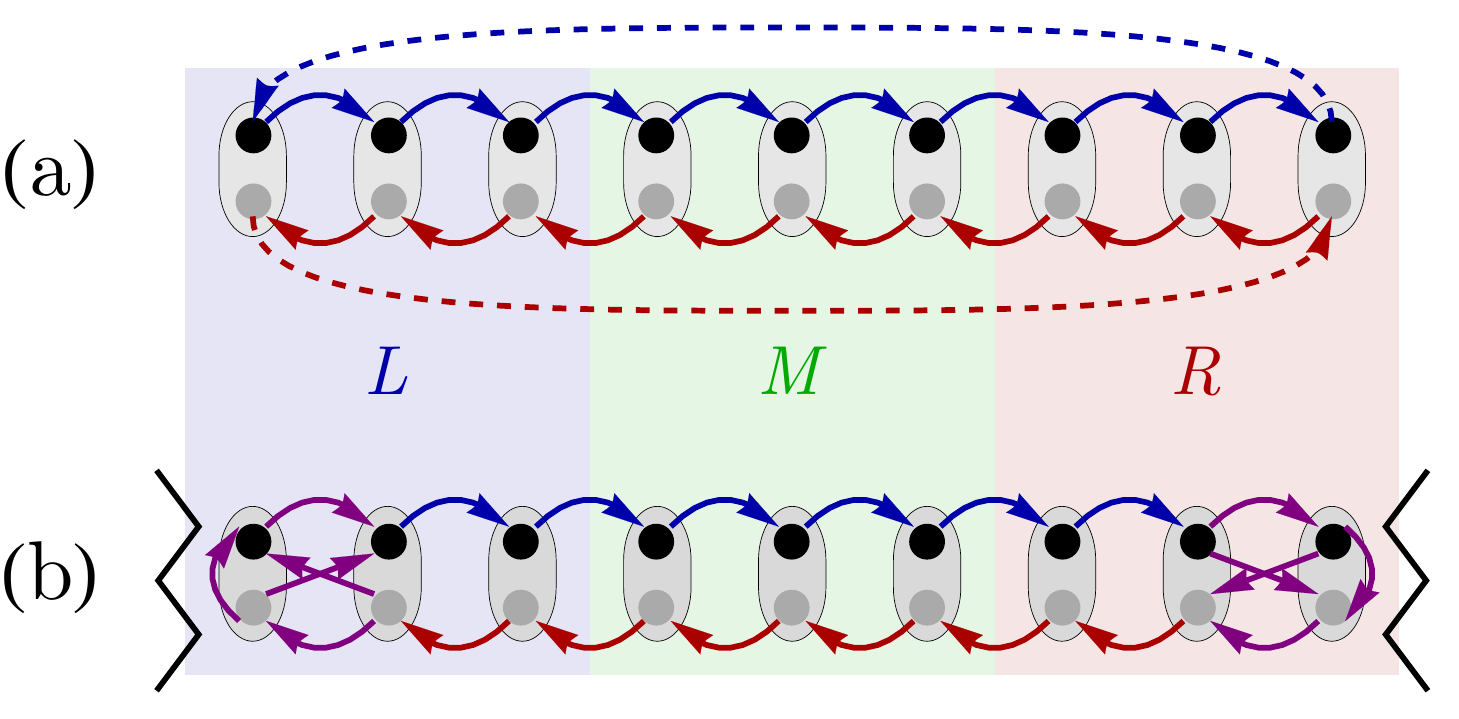}
\caption{Action of the general chiral unitaries $U_c(T/2)$ and $U_o(T/2)$ within the left edge ($L$), middle ($M$) and right edge ($R$) regions of a 1D chain. Sublattices are shown as black and gray points. (a) The unitary action of the closed system $U_c(T/2)$ acts on each sublattice independently. (b) In the open system, the unitary $U_o(T/2)$ acts on each sublattice independently in the bulk but couples the two sublattices in the edge regions. \label{fig:bulk_boundary}}
\end{figure}

We identify the three relevant spatial regions (left edge, middle, and right edge) as $L$, $M$ and $R$, respectively, as shown in Fig.~\ref{fig:bulk_boundary}. The bulk region $M$ should be defined far enough away from the edges (i.e. larger than the Lieb-Robinson length away) that $U_c(t)$ and $U_o(t)$ act identically within this region. For concreteness, a chain of length $N$ can be split into the regions
\beq
\begin{array}{cccc}
L :&& x\leq N/3\\
M :&& N/3<x\leq 2N/3\\
R :&& x>2N/3,
\end{array}
\eeq
rounding the fractions if necessary.

We recall that halfway through a chiral-symmetric drive, the unitary operator $U_c(T/2)$ takes a block diagonal form (see Eq.~\eqref{eq:block_diagonal}), indicating that the two sublattices become decoupled. This motivated the notion of chiral flow, which we defined in Eq.~\eqref{eq:chiral_flow}. In the open system, however, $U_o(T/2)$ will not in general take this block-diagonal form. Instead, at the edges of the system there may be coupling between the two sublattices, as we found for the model drive and as illustrated schematically in Fig.~\ref{fig:bulk_boundary}. However, within the bulk region $M$, both closed- and open-system drives are identical. In this way, we can calculate the chiral flow invariant $\nu[U]$ in the bulk even for the open system.

We now use the properties of chiral flow to relate the bulk behaviour to the boundary behaviour of the open system at $t=T/2$. As noted in Sec.~\ref{sec:unitary_flow}, chiral flow builds on the notion of unitary flow from Ref.~\onlinecite{KitaevAnyons2006}, and inherits many of its properties. Importantly, the unitary flow invariant is independent of location, and may be calculated across any imaginary cut in the 1D system. In turn, this implies that unitary flow is constant and conserved throughout the system. 

In the bulk at $t=T/2$, chiral flow is similarly constant and conserved. However, at the edge region $R$, unitary flow corresponding to one sublattice flows from region $M$ to region $R$, while flow corresponding to the other sublattice flows from $R$ to $L$. Since unitary flow is conserved, there must be flow between sublattices at some point within the edge region. This is the coupling between sublattices that is shown schematically in Fig.~\ref{fig:bulk_boundary}. We show below that this inter-sublattice flow at the edges is \emph{exactly} equal to the chiral flow in the bulk.

We write the open-system half-period unitary $U_o(T/2)$ using the shorthand $U$ and consider the trivial trace
\beq
0&=&\mathrm{Tr}\left[U^{-1}P_{R,A}U-P_{R,A}\right],
\eeq
where $P_{R,A}=P_RP_A$ is a projector onto sublattice A in region $R$. This trace vanishes because, for an open system, we can use the cyclic property of the trace to bring $U$ next to $U^{-1}$ and replace $UU^{-1}=\id$. Defining the complementary projector $\bar{P}_{R,A}=\id-{P}_{R,A}$, we can rewrite the above trace as
\beq
0&=&\mathrm{Tr}\left[U^{-1}P_{R,A}U\bar{P}_{R,A}-U^{-1}\bar{P}_{R,A}UP_{R,A}\right].
\eeq
The complementary projector can be expanded as a sum over all other sublattices and regions,
\beq
\bar{P}_{R,A}&=&P_{L,A}+P_{L,B}+P_{M,A}+P_{M,B}+P_{R,B}.
\eeq
However, assuming the unitary has some finite strict localisation length, there can be no overlap between $U^{-1}P_RU$ and $P_L$, and so we can ignore $P_L$ in the complementary projector. In addition, in the bulk the unitary preserves chiral symmetry, and so $U^{-1}P_{R,A}U$ can have overlap with $P_{M,A}$ but not with $P_{M,B}$. Expanding the remaining projectors, we find
\beq
0&=&\mathrm{Tr}\left[U^{-1}P_{R,A}UP_{M,A}-U^{-1}P_{M,A}UP_{R,A}\right]\nonumber\\
&&+\mathrm{Tr}\left[U^{-1}P_{R,A}UP_{R,B}-U^{-1}P_{R,B}UP_{R,A}\right].
\eeq
Referring back to Sec.~\ref{sec:real_space_invariant}, we identify the first trace as the chiral flow invariant $\nu[U]$, measured across the boundary between regions $M$ and $R$. Moving the second trace to the other side of the equation, we identify it as an edge invariant, equal to the chiral flow, which captures the flow between sublattices within the region $R$,
\begin{equation}
\nu^R_{\rm edge}[U]=\mathrm{Tr}\left[U^{-1}P_{R,B}UP_{R,A}-U^{-1}P_{R,A}UP_{R,B}\right].\label{eq:edge_invariant_half1}
\end{equation}
This can be written equivalently as
\begin{equation}
\nu^R_{\rm edge}[U]=\mathrm{Tr}\left[U^{-1}P_{B}UP_AP_{R}-U^{-1}P_{A}UP_BP_{R}\right],\label{eq:edge_invariant_half}
\end{equation}
where we have dropped two projectors onto $R$, which are unnecessary because any flow into region $M$ must conserve sublattice. We can define a similar edge invariant at the left edge, which is equal in magnitude but opposite in direction to $\nu^R_{\rm edge}[U]$.

Overall, we see that the bulk chiral flow invariant (Eq.~\eqref{eq:chiral_flow}) is equal to the half-period edge invariant (Eq.~\eqref{eq:edge_invariant_half}), i.e. that 
\beq
\nu[U_c(T/2)]=\nu[U_o(T/2)]=\nu_{\rm edge}^{R}[U_o(T/2)].
\eeq
This is the first step in our derivation of bulk-edge correspondence. 

\subsection{Bulk-Edge Correspondence at $t=T$\label{sec:bulk-edge_T}}
To complete the derivation, we now show that the half-period edge invariant is equal to the number of protected edge modes present at the right-hand edge at the end of the evolution, $n_\pi^{R}$.

To do this, we rewrite our expression for $\nu^R_{\rm edge}[U]$ in terms of $C$ by substituting Eq.~\eqref{eq:pab_chiral} into Eq.~\eqref{eq:edge_invariant_half}. Writing out $U=U_o(T/2)$ in full, this gives
\begin{equation}
\nu_{\rm edge}^R[U]=\frac{1}{2}\bigg[\mathrm{Tr}\left[CP_R\right]-\mathrm{Tr}\left[U^{-1}_o(T/2)CU_o(T/2)P_R\right]\bigg].
\end{equation}
However, Eq.~\eqref{eq:half_period_full_period_relation} gives a relation between a generic half-period unitary and the corresponding full-period unitary, which we can use to rewrite the expression above as
\beq
\nu_{\rm edge}^R[U]&=&\frac{1}{2}\bigg[\mathrm{Tr}\left[CP_R\right]-\mathrm{Tr}\left[U^{-1}_o(T)CP_R\right]\bigg]\nonumber\\
&=&-\frac{1}{2}\mathrm{Tr}\left[C\left(U^{-1}_o(T)-\id\right) P_R\right],
\eeq
where in the second line we have used that $C$ commutes with $P_R$ and grouped together the expressions under the trace. Finally, we identify the expression above as $n_\pi^R$ from Eq.~\eqref{eq:edge_invariant}, noting that either $U_o(T)$ or $U_o^{-1}(T)$ may be used to count edge modes at $\epsilon=\pi$.

Overall, we have shown that chiral flow in the bulk at $t=T/2$ leads to chiral symmetry-breaking flow at the boundary, which in turn generates edge modes at $\epsilon=\pi$ at the end of the evolution. This bulk-boundary correspondence is summarised by the three equal invariants
\beq
\nu[U_{c/o}(T/2)]=\nu_{\rm edge}^{R}[U_o(T/2)]=n_\pi^{R}[U_o(T)].
\eeq

\section{Conclusion\label{sec:conclusion}}
In this work, we have introduced chiral flow (Eq.~\eqref{eq:chiral_flow}) as a physically motivated, locally computable bulk invariant, which describes the topological properties of unitary evolutions with chiral symmetry. While the invariant itself is defined only for unitary loop evolutions, we argued in Sec.~\ref{sec:Prelim} and in Appendix~\ref{app:loop_from_unitary} that \emph{any} chiral evolution (with a gap at $\epsilon=\pi$) is related to a characteristic unitary loop. In this way, chiral flow provides a topological characterisation of any (gapped and non-interacting) driven system belonging to class~AIII. This invariant is an improvement on previous invariants, discussed in Sec.~\ref{sec:intro}, in that it applies to systems with disorder and is locally computable. In addition, it has the intuitive physical interpretation of describing the unitary flow \cite{KitaevAnyons2006} on each sublattice at the half-period point.

We went on to derive an explicit bulk-boundary correspondence which relates the chiral flow to the number of protected dynamical edge modes present at the end of the evolution. To do this, we first introduced an edge invariant (Eq.~\eqref{eq:edge_invariant_half}) which quantifies the chiral-symmetry-breaking flow that arises at a boundary at the midpoint of the evolution. This was found to be exactly equal to the chiral flow invariant in the bulk. It is interesting to note that the behaviour of a chiral drive at $t=T/2$ offers much more information about its topological properties than its behaviour at $t=T$.

Finally, we equated this half-period edge invariant to the full-period edge invariant (Eq.~\eqref{eq:edge_invariant}), which directly counts the number of protected edge modes at $\epsilon=\pi$ at the end of the evolution. In this way, our work provides the first explicit bulk-edge correspondence for one-dimensional Floquet systems with chiral symmetry. In passing, we note that the full-period edge invariant we introduced may be used to count the number of modes at $\epsilon=\pi$ in general, and is applicable beyond unitary loop evolutions.

Our work raises a number of interesting open questions. First, Floquet system in class~AIII have been shown to host nontrivial topological phases in all odd (spatial) dimensions, but have thus far only been studied in cases with translational symmetry or in one dimension. In future work, we will extend the notion of chiral flow, and the associated bulk-boundary correspondence, to the higher dimensional case. In the process, we hope to provide insight into the boundary behaviour of Floquet topological phases in even dimensions. Our work also extends Kitaev's notion of unitary flow \cite{KitaevAnyons2006} to systems with chiral symmetry. It would be interesting to study whether this quantity can be similarly extended to the other symmetry classes in the 10-fold way. Finally, an information-theoretic extension of the notion of flow was applied to many-body unitary evolutions in Ref.~\onlinecite{GrossIndex2012}, in the context of quantum cellular automata. The resulting topological invariant underpins the classification of interacting Floquet topological phases in two and three dimensions introduced in Refs.~\onlinecite{PoChiral2016,HarperFloquet2017,ReissInteracting2018}. An extension of this many-body invariant to systems with chiral symmetry, and indeed in other symmetry classes and dimensions, remains an interesting avenue for future research.

\begin{acknowledgments}
X.~L, F.~H., and R.~R. acknowledge support from the NSF under CAREER DMR-1455368 and the Alfred P. Sloan foundation.
\end{acknowledgments}

\appendix
\section{Extension to Exponentially Decaying Unitaries\label{app:exponentially_decaying}}
In the main text, the unitary operators we considered were assumed to be strictly local with some localization length $\ell$, i.e., we assumed that $U_{jk}=0$ for $|j-k|>\ell$ (where $j$ and $k$ label unit cell positions). In this appendix, we extend our results to the more general definition of locality in which matrix element magnitudes decay exponentially with distance. Specifically, we will assume that for large enough $|j-k|$, the unitary operator satisfies
\beq
\left|U_{jk}\right|\leq Ce^{-\left|j-k\right|/\ell},\label{eq:exp_loc}
\eeq
for some positive constant $C$ and localization length $\ell$. If we evolve a local Hamiltonian in time, the unitary time-evolution operator will generically take this form \cite{GrafBulk2018}.

\begin{figure}[t]
\includegraphics[scale=0.55]{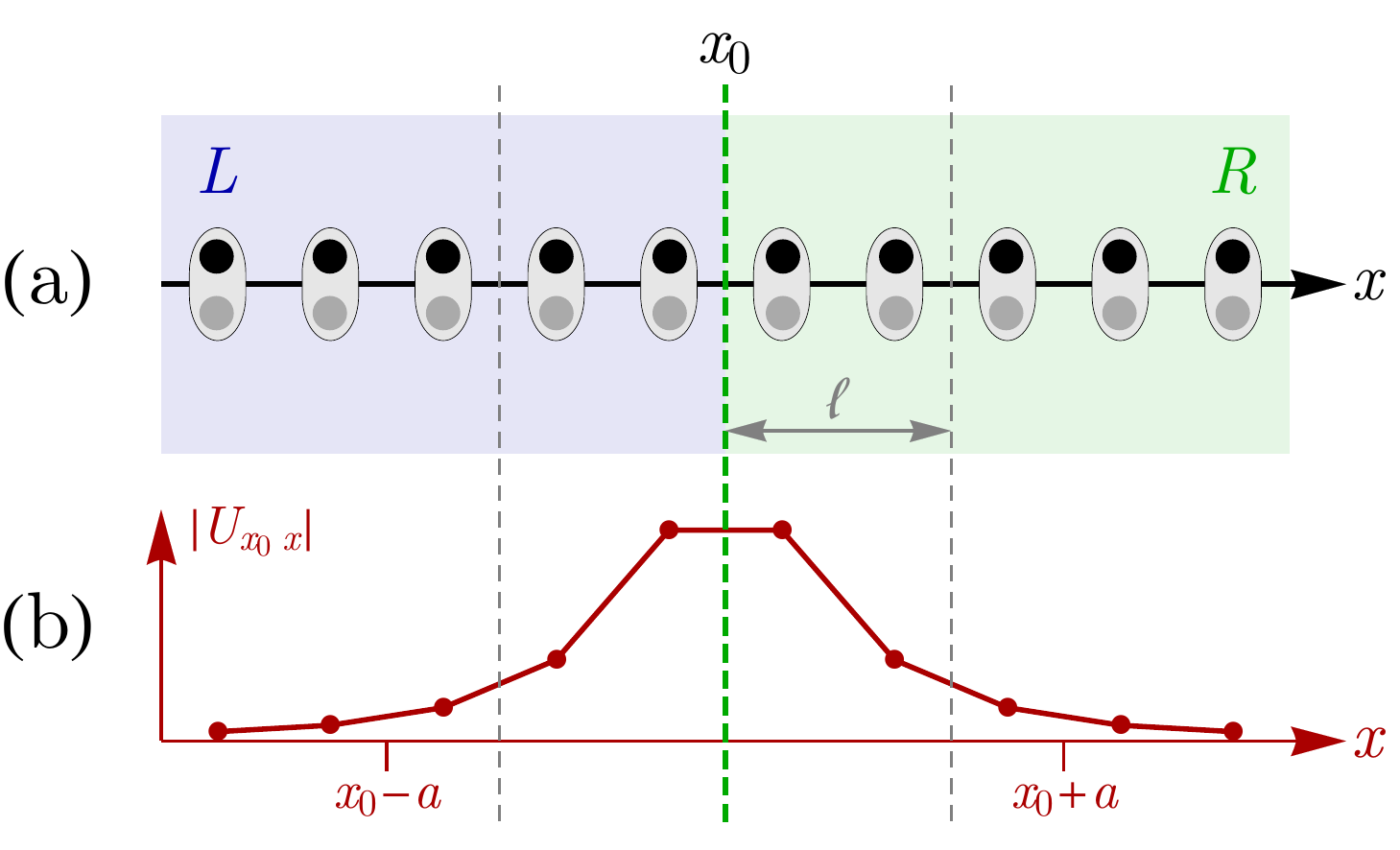}
\caption{Calculation of the chiral flow invariant for a local unitary satisfying Eq.~\eqref{eq:exp_loc}. (a) We split the region near a cut at $x=x_0$ into left ($L$) and right ($R$) pieces. (b) Matrix elements which connect sites across the cut are bounded by an exponentially decaying envelope function. By truncating projectors to sites within a distance $a$ from the cut, we neglect contributions which are of size $O(e^{-a/\ell})$. By taking $a$ much larger than $\ell$ these errors can be made exponentially small.\label{fig:truncation} }
\end{figure}

\subsection{Bulk Invariants}
We now study how this looser definition of locality affects the definition and calculation of invariants introduced in the main text. We recall that the bulk chiral flow invariant (Eq.~\eqref{eq:chiral_flow}) may be written for a formally infinite system as
\begin{equation}
\nu[U]=\frac{1}{2}\bigg[\textrm{\textrm{Tr}}\left(CU^{-1} P_RUP_L\right)-\textrm{\textrm{Tr}}\left(CU^{-1} P_LUP_R\right)\bigg],\label{eq:chiral_flow_appendix}
\end{equation}
where $U=U(T/2)$ and $P_L$ and $P_R$ are projectors onto the semi-infinite regions $x\leq x_0$ and $x>x_0$, respectively. For the infinite system, this equation continues to give a well-defined quantised chiral flow index, even for exponentially localised unitaries satisfying Eq.~\eqref{eq:exp_loc} \cite{KitaevAnyons2006}. For practical purposes, however, infinite system sizes cannot be achieved, and we must necessarily consider a finite system with projectors truncated to a finite region around the cut at $x_0$. 

In the main text, the unitary operators we considered were strictly local, and we could truncate to the region $\left[x_0-\ell,x_0+\ell\right]$ without changing the value of $\nu[U]$. In addition, $\nu[U]$ could be calculated using either the closed-system unitary $U_c(T/2)$ or the open-system unitary $U_o(T/2)$, since these had identical action within the truncated region (as long as the edges were far enough away from $x_0$). In the current case, the open and closed system unitaries will have actions which differ at $x_0$, even if only by an exponentially small amount. In addition, truncation to a region around a cut at $x_0$ will introduce other (exponentially small) errors, and so too will having a finite system size. All of these sources of error will need to be taken into account and quantified. 

First, we consider a closed system with $N$ sites in total (which we write as $c[N]$) and attempt to calculate the bulk invariant using a truncation to the region $\left[x_0-a,x_0+a\right]$, with $a\leq N/2$.\footnote{For a finite system, we require $a\leq N/2$ so that only the flow across the cut is measured, and not contributions which pass around the `back' of the system.} Explicitly, we set $x_0=0$ and define
\begin{equation}
\nu_{c[N],a}[U]=\frac{1}{2}\bigg[\textrm{\textrm{Tr}}\left(CU^{-1} P_R^a UP_L^a \right)-\textrm{\textrm{Tr}}\left(CU^{-1} P_L^a UP_R^a \right)\bigg],\label{eq:bulk_invariant_trunc}
\end{equation}
where $P_L^a$ projects onto the range $[-a,0)$, $P_R^a$ projects onto the range $[0,a)$, and $U$ is shorthand for the closed-system unitary $U_{c[N]}(T/2)$. It is clear that
\beq
\lim_{a\to\infty}\lim_{N\to\infty}\nu_{c[N],a}[U]&=&\nu[U],
\eeq
as this recovers Eq.~\eqref{eq:chiral_flow_appendix} which gives the exact, quantised invariant. 

For large but finite values of $a$ and $N$, this expression will neglect contributions from the unitary operator with magnitudes less than or equal to $O(e^{-a/\ell})$ and $O(e^{-N/(2\ell)})$, respectively. Since $a\leq N/2$, the errors due to finite system size will be smaller than those due to truncation, and so we can safely ignore them. The total error in the calculated value of $\nu_{c[N],a}[U]$ is bounded by the sum of all neglected terms, and so overall we expect
\beq
\nu_{c[N],a}[U]&=&\nu[U]+O\left(e^{-a/\ell}\right).
\eeq
In this way, by taking the truncation length $a\gg\ell$ (and increasing the system size correspondingly), it is possible to calculate the chiral flow invariant to arbitrary accuracy. This idea is shown schematically in Fig.~\ref{fig:truncation}.

We now consider a finite open system $o[N]$, which has $N$ sites labelled from $-N/2$ to $N/2-1$. We again try to calculate the bulk invariant using a truncation to the region $[-a,a]$, and this time define
\begin{equation}
\nu_{o[N],a}[U]=\frac{1}{2}\bigg[\textrm{\textrm{Tr}}\left(CU^{-1} P_R^a UP_L^a \right)-\textrm{\textrm{Tr}}\left(CU^{-1} P_L^a UP_R^a \right)\bigg],\label{eq:bulk_invariant_trunc_open}
\end{equation}
where definitions are as before except $U$ is now shorthand for the open-system unitary $U_{o[N]}(T/2)$. As for the closed system, we can take the limit $N\to\infty$ followed by the limit $a\to\infty$ to find
\beq
\lim_{a\to\infty}\lim_{N\to\infty}\nu_{o[N],a}[U]&=&\nu[U],
\eeq
which again recovers Eq.~\eqref{eq:chiral_flow_appendix} (since boundary conditions are negligible in the infinite system limit). At finite system sizes, there are errors due to the truncation and errors due the finite size. The scaling follows as before, and we find
\beq
\nu_{o[N],a}[U]&=&\nu[U]+O\left(e^{-a/\ell}\right).
\eeq
In this way, although calculations of the chiral flow invariant in the closed system and in the open system may be different, both values tend towards the same quantised value in the limit of infinite system size and infinite truncation region. At finite sizes the calculated values differ from the true value by errors of size $O(e^{-a/\ell})$, and correspondingly may differ from each other by a similar amount.

\subsection{Edge Invariants}
\begin{figure}[t]
\includegraphics[scale=0.55]{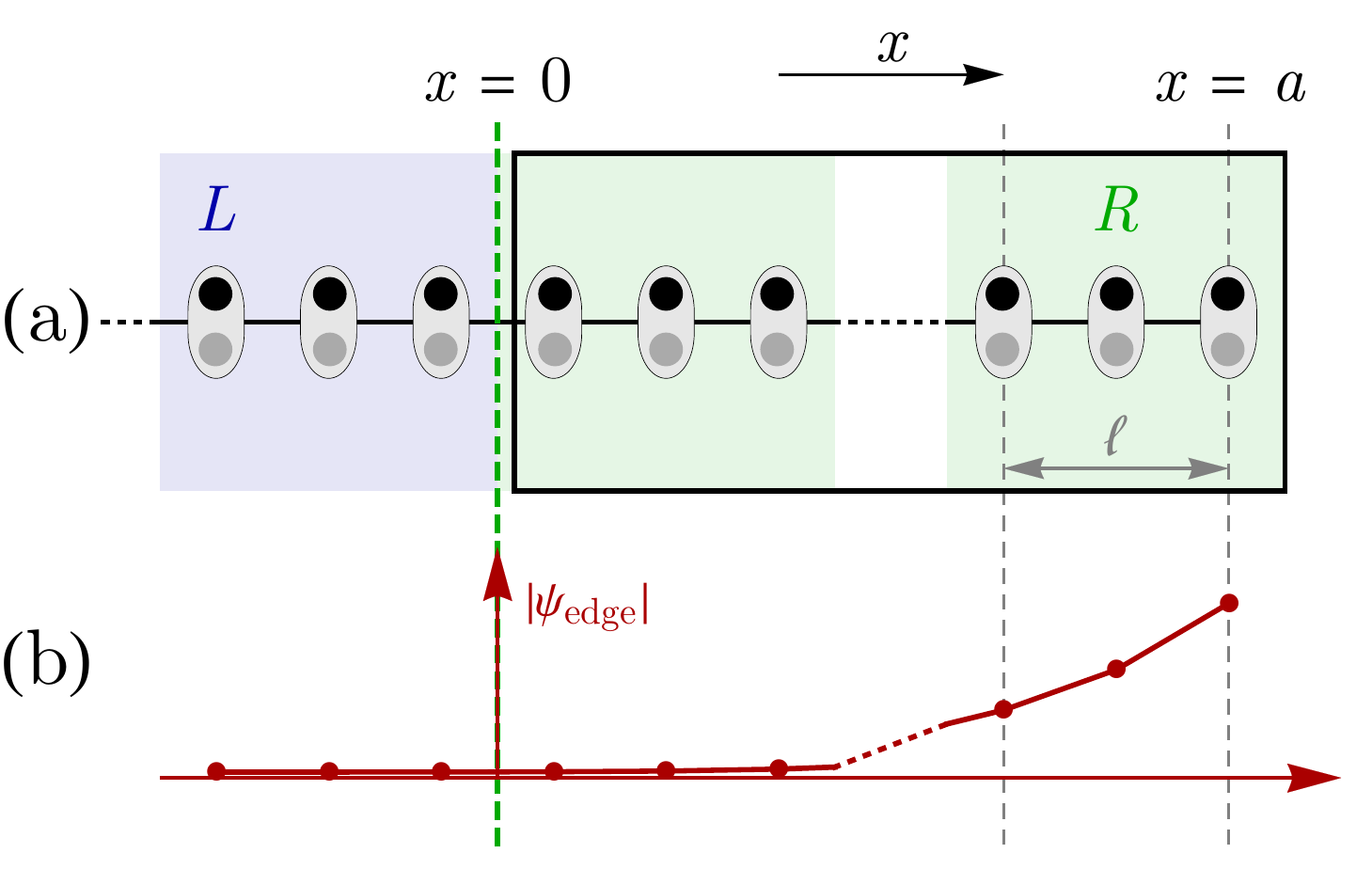}
\caption{Semi-infinite system used in the construction of truncated edge invariants. (a) Boxed region indicates the region where $P_R^a$ is nonzero, extending from $x=0$ to $x=a$. To reduce truncation errors, this should be much larger than the Lieb-Robinson length of the unitary $\ell$. (b) Edge modes decay exponentially away from $x=0$. Truncating to the boxed region means contributions of size $O(e^{-a/\ell})$ are neglected.\label{fig:edge_truncation} }
\end{figure}

We now consider the effects of the new definition of locality on the edge invariants defined in the main text. To aid the discussion, we formally consider a semi-infinite system extending from negative infinity to $x=a$, as shown in Figure.~\ref{fig:edge_truncation}. Recalling Eq.~\eqref{eq:edge_invariant_half}, we write the edge invariant for the semi-infinite system as 
\begin{equation}
\nu^R_{\mathrm{edge},a}[U]=\mathrm{Tr}\left[U^{-1}P_{B}UP_AP_{R}^a-U^{-1}P_{A}UP_BP_{R}^a\right].\label{eq:edge_invariant_trunc}
\end{equation}
where $U=U_o(T/2)$ is the open-system unitary and $P_R^a$ is a projector onto the right-hand edge region of the system, the interval $[0,a]$. For a strictly local unitary (as considered in the main text), the expression above gives an exact, quantised value, as long as $a>\ell$. For exponentially decaying unitary operators satisfying Eq.~\eqref{eq:exp_loc}, however, the definition of $R$ amounts to a truncation. In this case, the `edge region' should formally include exponentially small contributions (set by the length scale $\ell$) even on sites outside of the range $[0,a]$. The truncation in Eq.~\eqref{eq:edge_invariant_trunc} neglects these contributions of size $O(e^{-a/\ell})$, leading to a total error (bounded by a sum of these pieces), which is also of size $O(e^{-a/\ell})$. In the thermodynamic limit we take $a\to\infty$ and find
\beq
\lim_{a\to\infty}\left[\nu^R_{\mathrm{edge},a}[U]\right]&=&\nu^R_{\mathrm{edge}}[U].
\eeq
We will verify that this is indeed the true, quantised edge invariant below.

We first use a similar method as in the main text to show that the bulk and edge invariants are equal in the thermodynamic limit. Starting from the trivial trace
\beq
0&=&\mathrm{Tr}\left[U^{-1}P^a_{R,A}U-P^a_{R,A}\right],
\eeq
we insert the complementary projector $\overline{P^a_{R,A}}=P^a_{R,B}+P_{L,A}+P_{L,B}$, where $P_L$ projects onto the region $x\leq0$, and find
\beq
0&=&\mathrm{Tr}\left[U^{-1}P^a_{R,A}U\overline{P^a_{R,A}}-P^a_{R,A}\overline{P^a_{R,A}}\right]\nonumber\\
&=&\tilde{\nu}^R_{a}[U]-\nu^R_{\mathrm{edge},a}[U],
\eeq
where
\beq
{\nu}^R_{\mathrm{edge},a}[U]&=&\mathrm{Tr}\left[U^{-1}\left(P_{R,B}^a+P_{L,B}\right)UP^a_{R,A}\right.\nonumber\\
&&\left.-U^{-1}P^a_{R,A}U\left(P_{R,B}^a+P_{L,B}\right)\right]
\eeq
recovers Eq.~\eqref{eq:edge_invariant_trunc} and
\begin{equation}
\tilde{\nu}^R_{a}[U]=\mathrm{Tr}\left[U^{-1}P^a_{R,A}UP_{L,A}-U^{-1}P_{L,A}UP^a_{R,A}\right]
\end{equation}
is equivalent to Eq.~\eqref{eq:bulk_invariant_trunc} up to corrections of size $O(e^{-a/\ell})$ (due to the fact that the system is now semi-infinite). In this way, in the thermodynamic limit we find
\beq
\lim_{a\to\infty}\left[\nu^R_{\mathrm{edge},a}[U]\right]=\lim_{a\to\infty}\left[\tilde{\nu}^R_{a}[U]\right]=\nu[U],
\eeq
which recovers the quantised bulk invariant. In this way, we have defined truncated edge and bulk invariants at $t=T/2$ which are equal to each other and to their values in the thermodynamic limit up to exponentially small corrections. By taking the truncation region $a\gg\ell$, the corrections can be made exponentially small. This is illustrated schematically in Fig.~\ref{fig:edge_truncation}.

Finally, we define the invariant which counts the number of edge modes at $t=T$ in a similar way to Eq.~\eqref{eq:edge_invariant}. In the semi-infinite system, however, there is only one edge, and we can formally define the exact (quantised) number of edge modes as
\begin{equation}
n_{\pi}^R[U]=-\frac{1}{2}\mathrm{Tr}\left[C\left(U_o(T)-\id\right)\right],
\end{equation}
where $U_o(T)$ is obtained by removing terms from the generating Hamiltonian that connect sites across the boundary at $x=a$. In contrast to Eq.~\eqref{eq:edge_invariant}, there is no projector $P_R$, which previously served to remove any contributions from the left edge. This expression formally gives the exact number of edge modes, even for exponentially decaying unitaries satisfying Eq.~\eqref{eq:exp_loc}.

For any finite system, however, we must introduce a truncation, and so we define
\begin{equation}
n_{\pi,a}^R[U]=-\frac{1}{2}\mathrm{Tr}\left[C\left(U_o(T)-\id\right) P_R^a\right],
\end{equation}
where $P_R^a$ again projects onto the region between $x=0$ and $x=a$. This expression differs from the exact value by an error with size $O(e^{-a/\ell})$, as it neglects the exponential tails of the edge modes that permeate beyond $x=0$. However, these corrections are again of size $O(e^{-a/\ell})$, and can be made arbitrarily small by taking $a$ much larger than $\ell$. In this way,
\beq
\lim_{a\to\infty}\left[n_{\pi,a}^R[U]\right]&=&n_{\pi}^R[U].
\eeq
As in the main text, we can use Eq.~\eqref{eq:half_period_full_period_relation} to show that expressions for $n_{\pi,a}^R[U]$ and ${\nu}^R_{\mathrm{edge},a}[U]$ are equivalent. This argument is very similar to that given in Sec.~\ref{sec:bulk-edge_T}, and so we do not reproduce it here.

Overall, we find that bulk and edge invariants can be defined even for unitary operators satisfying the looser definition of locality given in Eq.~\ref{eq:exp_loc}. While these invariants take quantised values only in the limit of infinite system size, any realistic measurement necessarily requires truncation, which will introduce exponentially small corrections. However, by taking the truncation region to be much larger than the Lieb-Robinson length of the unitary operator, these errors can be made arbitrarily small.

\section{Obtaining a Unitary Loop from a General Unitary Evolution\label{app:loop_from_unitary}}
In the main text, we mostly worked with unitary loop evolutions, which satisfy $U_c(T)=\id$. Most unitary evolutions, however, will not satisfy this property. In these cases, as motivated in Sec.~\ref{sec:Prelim}, we may construct a unitary loop from the evolution which captures its inherently dynamical component. In this appendix, we outline this construction in more detail.

We consider an arbitrary closed-system evolution with chiral symmetry which, at $t=T$, has a gap in the quasienergy spectrum at $\epsilon=\pi$. [In order for there to be protected dynamical edge modes, we require a gap at $\epsilon=\pi$ in the closed system, and so we only consider this case here]. We can then define a Floquet Hamiltonian corresponding to this gap as
\beq
H_{F}&=&\frac{i}{T}\log_{\pi}U_c(T),
\eeq
where $\log_{\pi}$ is the complex logarithm defined as 
\beq
\log_{\pi}(e^{i\psi})=i\psi
\eeq
for
\beq
-\pi<\psi<\pi.
\eeq
Explicitly, if we express the full unitary evolution $U_c(T)$ in its eigenbasis,
\beq
U_c(T)=\sum_j \lambda_j \ket{\Psi_j}\!\bra{\Psi_j},
\eeq
then the Floquet Hamiltonian may be written
\beq
H_{F}&=&\frac{i}{T}\sum_j \log_{\pi}\left(\lambda_j\right) \ket{\Psi_j}\!\bra{\Psi_j}.
\eeq
As shown in Ref.~\onlinecite{GrafBulk2018}, a Floquet Hamiltonian defined in this way is local (in that the magnitudes of its matrix elements decay exponentially with distance). In addition, since the underlying evolution is chiral symmetric, the Floquet Hamiltonian satisfies 
\beq
CH_FC^{-1}&=&-H_F.
\eeq

We can deform the full unitary evolution $U_c(t)$ into a unitary loop followed by an evolution with $H_F$. First, we define a (chiral-symmetric) unitary loop through the generating Hamiltonian
\beq
H_L(t)&=&\left\{\renewcommand\arraystretch{1.2}\begin{array}{ccc}
-2H_{F} && 0\leq t<\frac{1}{4} T\\
2H(2(t-\frac{1}{4}T)) && \frac{1}{4} T \leq t< \frac{3}{4} T\\
-2H_F&& \frac{3}{4} T \leq t<  T,
\end{array}\right.,\label{eq:Loop_CH}
\eeq
where $H(t)$ is the original generating Hamiltonian for $U_c(t)$. It may be verified that the evolution
\beq
V_c(t)&=&\mathcal{T}\exp\left[-i\int_0^tH_L(t')\,\dd t'\right]
\eeq
satisfies $V_c(T)=\id$. To recover $U_c(T)$, we can evolve with $H_F$ for time $T/2$ before and after the evolution with $H_L(t)$ and note that
\begin{equation}
e^{-iH_FT/2}V_c(T)e^{-iH_FT/2}=e^{-iH_FT}\equiv U_c(T).
\end{equation}
This complete evolution has chiral symmetry and is homotopically connected to the original evolution $U_c(t)$ \cite{RoyPeriodic2017}. Dynamical edge modes can only arise during the evolution with $H_L(t)$, as it is only in this part of the evolution that the gap at $\epsilon=\pi$ can close. In this way, the dynamical properties of $U_c(t)$ are equivalent to the dynamical properties of the loop evolution $V_c(t)$.

The loop evolution $V_c(t)$ can be used directly in the calculation of the bulk chiral flow invariant in Eq.~\eqref{eq:chiral_flow}. For the edge invariants, we require the corresponding open system evolution, $V_o(t)$. This can be obtained by truncating the loop generating Hamiltonian $H_L(t)$ in Eq.~\eqref{eq:Loop_CH} by removing terms which connect sites across the boundary. Evolution with this open-system Hamiltonian then yields $V_o(t)$. This can be used in the calculation of the half-period edge invariant in Eq.~\eqref{eq:edge_invariant_half} and in the calculation of the number of edge modes at $t=T$ in Eq.~\eqref{eq:edge_invariant}.

\end{document}